\documentclass[a4paper,11pt]{article}
\pdfoutput=1 

\usepackage{jheppub} 

\usepackage[T1]{fontenc} 

\usepackage{amsmath}
\usepackage{amssymb}
\usepackage{graphicx}
\usepackage{xcolor}
\usepackage{multirow}
\usepackage{float}
\usepackage{makecell}
\usepackage{longtable}
\usepackage{subcaption} 
\usepackage{float}

\preprint{USTC-ICTS/PCFT-26-16}

\title{Wilson Surface One-Point Functions: A Case Study}

\author[a]{Long-Fu~Zhang,}
\author[a, b]{Jun-Bao~Wu}
\affiliation[a]{Center for Joint Quantum Studies and Department of Physics, School of Science, Tianjin University, 135 Yaguan Road, Tianjin 300350, P. R. China }
\affiliation[b]{Peng Huanwu Center for Fundamental Theory, 96 Jinzhai Road, Hefei, Anhui 230026, P. R. China}

\abstract{We compute holographic one-point functions for Wilson surfaces in the case of a toroidal surface operator. Compared to the cases of a planar or spherical surface operator, these one-point functions exhibit a more intricate dependence on the shape and position of both the surface and the local operators. Averaging over the moduli space of membranes dual to the surface operator plays a key role in the computations. We obtain both analytical and numerical results. The case of a cylindrical surface operator is also studied.}

\emailAdd{zhanglongfu@tju.edu.cn}
\emailAdd{junbao.wu@tju.edu.cn}

\begin{document} 
\maketitle
\flushbottom
\section{Introduction}

Many aspects of eleven-dimensional M-theory~\cite{Townsend:1995kk, Witten:1995ex} are still mysterious. The low energy effective theory (LEET) is the eleven-dimensional supergravity theory. The BPS branes in theory include M2-branes and M5-branes. The LEET of multiplet M2-branes was established as super Chern-Simons theory with gauge group $U(N)\times U(N)$, having Chern-Simons levels $1$ and $-1$, respectively~\cite{Aharony:2008ug}. On the other hand, we know very little about the LEET of multiple M5-branes which is the six-dimensional $(2, 0)$ superconformal theory.

We still lack a Lagrangian description for non-abelian $(2,0)$ theories, partly due to the fact that their two-form field, $B_{\mu\nu}$,  has a self-dual field strength. One way to study this theory is to place it on 
$\mathbb{R}^{1, 4}\times S^1$
  to obtain a five-dimensional maximally supersymmetric Yang–Mills (MSYM) theory. This can be obtained from the string duality between M-theory and type IIA superstring theory: After compactifying on a circle, M5-branes become D4-branes in type IIA theory. More than a decade ago, it was proposed that the $5d$ MSYM theory with all instanton particles included already provides the complete non-perturbative description of the 
$6d$ $(2,0)$ theory~\cite{Douglas:2010iu, Lambert:2010iw}. Despite many efforts, whether this proposal is completely correct remains  debated.

Another approach to this six-dimensional $A_{N-1}$ $(2, 0)$ theory is via the AdS/CFT correspondence~\cite{Maldacena:1997re}. This $6d$ theory is dual to the M-theory on $AdS_7\times S^4$ background. When $N$ is large, the quantum effect in M-theory is suppressed, and supergravity  theory on $AdS_7\times S^4$ provides a good description of the field theory.

Non-local operators play a key role in the study of quantum field theory in general. In $6d$ 
$(2, 0)$ theories, Wilson surface operators are one of the most important non-local operators.  In Abelian $(2, 0)$ theory ($N=1$), the ordinal Wilson surface is defined as 
\begin{equation}
    W[S]=\exp \left( i\int_S P[B] \right),
\end{equation}
where $P[B]$ is the pullback of the two-form field $B$ to the surface $S$. The definition of (locally) BPS Wilson-surface includes the coupling to the five scalars $\Phi_I$
\begin{equation}
    W[S]_{\textrm{BPS}}=\exp\left(i \int_S(P[B]-n^I \Phi_I)\textrm{Vol}[S]\right). 
\end{equation}
Here, $n^I$ is a unit five-vector ($n^In^I=1$), which may depend on the point of $S$. For the non-Abelian case, such a definition becomes a formal one,  partly because defining the order along the surface is a hard task. Nevertheless, BPS conditions can be investigated.
Many Wilson surfaces with rigid supercharges were discovered in~\cite{Drukker:2020bes}.

The $AdS_7/CFT_6$ correspondence provides a powerful tool for studying Wilson surfaces. The surface operator in the fundamental representation of $A_{N-1}$ is dual to an M2-brane in $AdS_7\times S^4$~\cite{Maldacena:1998im}. Wilson surfaces in symmetric or anti-symmetric representations are well described by an M5-brane~\cite{Lunin:2007ab, Chen:2007ir}. For higher representations, the surface operator is dual to a bubbling geometry that is asymptotically $AdS_7\times S^4$~\cite{Lunin:2007ab, DHoker:2008lup, DHoker:2008rje}.

For some situations, the dual description of a Wilson surface is not simply a brane but a collection of branes. The reason is that each individual brane preserves less symmetries comparing with the Wilson surface. Only the moduli space of dual branes has the correct symmetry. A consequence of this fact is that when we compute correlation functions involving such Wilson-surfaces, we should compute the average of the result from an individual brane over the above moduli space. One of the examples of such cases is the M2-brane description of $1/8$-BPS toroidal Wilson surfaces~\cite{Drukker:2021vyx}. The string theory description of $1/4$-BPS latitude Zarembo loops~\cite{Zarembo:2002an} also has this kind of moduli space which provides zero modes in the computation of the worldsheet one-loop partition function~\cite{Medina-Rincon:2018wjs}.

When the surface operator is planar or spherical~\cite{Berenstein:1998ij, Corrado:1999pi, Chen:2007zzr}, the spacetime dependence of the Wilson surface one-point function is completely fixed by symmetries~\cite{Alday:2011pf}.  For Wilson surface with generic shape, such a correlator is a general function of this shape and the positions of the operators.  With this and the orbit average in mind, we compute 
the correlation functions of the above toroidal Wilson surface and a local chiral primary operator (CPO).

This paper is organized as following:  in section~\ref{fields}, we provide the setup of the Wilson surface one-point functions.  
In section~\ref{holography}, after reviewing the holographic dual of the toroidal/cylindrical Wilson surfaces and the CPO, we compute the Wilson-surface one-point functions for these two types of surface operators. Some details about spherical harmonics are put in the Appendix~\ref{harmonics}.


\section{Wilson surface one-point functions}\label{fields}
\paragraph{Six-dimensional $(2, 0)$ theories.}
The field-theoretic content of the six-dimensional \((2,0)\) theory includes a 2-form field \(B_{\mu\nu}\) with self-dual field strength, five scalars $\Phi_I$, and four fermions. 
This theory possesses sixteen supercharges and a global symmetry of \(SO(2,6) \times SO_R(5)\). Taking into account supersymmetries and superconformal supersymmetries, the global symmety is $OSp(8^*|4)$. 
The \(A_{N-1}\) \((2,0)\) theory serves as the LEET for \(N\) coincident M5-branes~\cite{Strominger:1995ac}. 
For \(N \geq 2\), a satisfactory Lagrangian description of this theory remains elusive, which makes its quantitative study particularly challenging. 
One viable approach is to compactify the theory on a circle \(S^1_R\). 
When the circle's radius \(R\) is sufficiently small, the compactification yields five-dimensional MSYM theory, which describes the dynamics of multiple D4-branes in type IIA superstring theory. Another approach to study the \(A_{N-1}\) \((2, 0)\) theory is via the hologrphaic description in terms of M-theory  on \(AdS_7\times S^4\) background. This is the focus of the current paper. 

\paragraph{Toroidal Wilson surfaces.}
In this paper, we focus on toroidal Wilson surfaces and their M-theory duals studied in~\cite{Drukker:2021vyx}. One special property of these surface operators is that their observables do not exhibit  divergences due to conformal anomalies.  These operators are along the flat tori in $\mathbb{R}^4\subset \mathbb{R}^6$. 
They are coupled with two scalar fields in the free version of the $(2, 0)$ theory. In other words, the polarization vector $n^I$ only has two non-zero components. According to the classification in~\cite{Drukker:2020bes}, they belong to sub-type-L of Lagrangian surfaces in $\mathbb{R}^4$, and this sub-type in turn belongs to the type-$\mathbb{H}$. They are $1/8$-BPS since they preserve two Poincar\'e supercharges and two conformal supercharges. 

The surface  is specified  by the following parameterization
\begin{align}
    x^1&=R_1 \cos\varphi_1\,,\,\, x^2=R_1 \sin\varphi_1\,,\nonumber\\
    x^3&=R_2 \cos\varphi_2\,, \,\, x^4=R_2\sin \varphi_2\,,\nonumber\\
    x^5&=x^6=0.
\end{align}
with $ R_i, i=1, 2$ constants and $0\leq \varphi_1, \varphi_2 \leq 2\pi$. Without loss of
generality, we assume that $R_1\ge R_2$.
The polarization vector is 
\begin{equation}\label{scalars}
    n^1=-\sin (\varphi_1+\varphi_2)\,, \,\,
    n^2=\cos (\varphi_1+\varphi_2)\,.
\end{equation}
We consider the above Wilson surfaces in the fundamental representation of the $A_{N-1}$ Lie algebra which is the algebra associated with the $(2, 0)$ theory. When the theory is compactified on a $S^1$, this algebra is the gauge algebra of the obtained five dimensional MSYM.  The surface operator is dual to a probe M2-brane in the M-theory on $AdS_7\times S^4$. The membrane solution will be reviewed in Subsection~\ref{membrane}.

\paragraph{Local operators.}
We will study the correlation functions between the Wilson surfaces and a local chiral primary operator $\mathcal{O}_\Delta$.
The CPO  belongs to the  $k$-th symmetric traceless representation of $SO_R(5)$~\cite{Aharony:1998rm, Minwalla:1998rp, Leigh:1998kt, Halyo:1998mc}. Its conformal dimension is $\Delta=2k, k\ge 2$, protected by supersymmetries.
\section{Holographic computations}\label{holography}

\subsection{Background fields in the bulk}
$(2, 0)$ theory with Lie algebra $A_{N-1}$ in the Euclidean space $\mathbb{R}^6$ is dual to M-theory on $(E)AdS_7 \times S^4$. The non-trivial background fields include the metric and the four-form flux, 
\begin{align}
    ds^2&=ds^2_{AdS_7}+\frac14 ds^2_{S^4}\,,
    \\
    F_4&=\frac38 d\text{Vol}_{S^4}\,.
\end{align}
Here $d\text{Vol}_{S^4}$ is the volume form of the unit $S^4$ and  we have set the radius of $AdS_7$ to be one. Then the AdS/CFT dictionary~\cite{Maldacena:1997re} gives the following relation between eleven-dimensional Planck length $l_p$ and $N$,
\begin{equation}
l_p=(8\pi N)^{-\frac13}.
\end{equation}
The four-form field strength $F_4$ is related to the corresponding three-form gauge potential $C_3$ by 
\begin{equation}
    F_4=dC_3.
\end{equation}
The metric on the unit $AdS_7$
is \footnote{the indices convention: $m,n...$ refer to $AdS_7 \times S^4$, $\mu ,\nu...$ refer to $AdS_7$, $\alpha,\beta...$ refer to $S^4$, $i,j...$ refer to the boundary of $AdS_7$ where the six-dimensional (2, 0) field theory lies, $a,b...$ refer to the worldvolume.}
\begin{equation}
    ds^2_{AdS_7}=\frac1{z^2}(dx^i dx^i+dz^2),
\end{equation}
in the Ponicar\'e coordinates. 
The metric on the unit $S^4$ is
\begin{equation}
    ds^2_{S^4}=d\theta^2+\sin^2 \theta d\phi^2+ \cos^2\theta (d\alpha^2+\sin^2\alpha d\beta^2)\,.
\end{equation}
The angular coordinates $\theta, \phi, \alpha, \beta $ satisfy $0\leq \theta\leq \frac\pi2, 0\leq  \alpha\leq \pi,0\leq \phi, \beta \leq 2 \pi$. They are related to the unit vector $\theta^I$ in $\mathbb{R}^5$ as 
\begin{align}
    \theta^1&= -\sin \theta \sin \phi\,,\\
    \theta^2&=\sin \theta \cos\phi\,,\\
    \theta^3&=\cos \theta \cos \alpha\,,\\
    \theta^4&=\cos \theta \sin\alpha \cos\beta\,,\\
    \theta^5&=\cos\theta \sin\alpha \sin \beta\,.
\end{align}

\subsection{Probe membrane solutions}\label{membrane}

We now review the membrane solutions in~\cite{Drukker:2021vyx} dual to the toroidal Wilson surfaces, though in different bulk and worldvolume coordinate systems. 
We choose the worldvolume coordinates as $\{\sigma^1, \sigma^2, \sigma^3\}=\{\varphi_1, \varphi_2, z\}$. 
Then the solutions are 
\begin{align}\label{eq:torus}
    x^1&=r_1(z) \cos \varphi_1\,,\,\,
    x^2=r_1(z) \sin \varphi_1\,,\nonumber\\
    x^3&=r_2(z)\cos\varphi_2\,,\,\, x^4=r_2(z)\sin\varphi_2\,,\nonumber\\
    x^5&=x^6=0,\nonumber\\
    \phi&=\varphi_1+\varphi_2\,,\,\, 
    \cos\theta=\frac{z^2}{2R_2^2}\,,\nonumber
    \\\alpha&=\alpha_0\,,\,\, \beta=\beta_0\,,
\end{align}\label{torus solution}
with 
\begin{equation}
    r_1(z)=\sqrt{R_1^2-\frac{z^2}{2}}\,,\,\, r_2(z)=\sqrt{R_2^2-\frac{z^2}{2}}\,
\end{equation}
and $\alpha_0, \beta_0$ are constants.  The brane solution in \cite{Drukker:2021vyx} is the solution with $\alpha_0=0$. Here we would like to remark that the dual of the toroidal surface operator should be  the \emph{collection} of membrane solutions with $\alpha_0\in [0, \pi], \beta_0\in [0, 2\pi]$. The reason is that the individual solution breaks the $SO(3)$ isometry of the $\{(n_3, n_4, n_5)\}$ space into $SO(2)$. Only the whole moduli space of the solutions preserves this symmetry~\footnote{In the terminology of~\cite{Drukker:2008zx}, the brane is smeared on this moduli space which is the coset $SO(3)/SO(2) \sim S^2$.}. Consequently, in the holographic computations  of  vacuum expectation values (vev's) of toroidal surface operators and Wilson surface one-point functions,   we should average over this moduli space. This is just the average over the orbit of the action of the above $SO(3)$ group. Such an orbit average appears in the holographic computations of correlators involving operators dual to probe string or brane solutions~\cite{Bajnok:2014sza, Yang:2021kot}.

The bosonic part of the membrane action is 
\begin{equation}\label{eq:m2action}
    S_{\mathrm{M2}}=T_{\mathrm{M2}} \int (d\mathrm{Vol}-P[C_3]).
\end{equation}
Here $d\mathrm{Vol}$ is the volume form of the worldvolme and $P[C_3]$ is the pullback of the potential  $C_3$ to the worldvolume. We also have 
\begin{equation}
T_{\mathrm{M2}}=\frac1{4\pi^2 l_p^3}=\frac{2N}\pi
\end{equation}
Taking into account the Legendre transformation, the on-shell action of an individual probe membrane is 
\begin{equation}
    S_{\text{ren.}}^{\text{tor.}}=-\pi N \left(\frac{R_1}{R_2}+\frac{R_2}{R_1}\right)\,.
\end{equation}
Keeping in mind the average over $S^2$, the large N limit of the Wilson surface vev is 
\begin{equation}
    \langle W(S) \rangle =\frac{1}{\text{Vol}(S^2)}\int d\mu (S^2) \exp (- S^{\text{tor.}}_{\text{ren.}}), 
\end{equation}
where $d\mu (S^2)=\sin \alpha_0 d\alpha_0 d\beta_0$, and $\text{Vol}(S^2)=\int d\mu (S^2)=4\pi$. For the case at hand, since $ S^{\text{tor.}}_{\text{ren.}}$ is independent of $\alpha_0, \beta_0$, we simply have 
\begin{equation}
    \begin{split}
         \langle W(S)\rangle &= \exp (- S^{\text{tor. }}_{\text{ren.}})\\
    &=\exp \left[\pi N\left( \frac{R_1 }{R_2}+\frac{R_2}{R_1}\right)\right].
    \end{split}
\end{equation}
Cylindrical Wilson surfaces are the limits of the toroidal Wilson surfaces with $R_1 \to \infty$ and $R_2$ fixed. The cylinder can be put at
\begin{align}
&    x^1=v, \, \, x^2=0,\nonumber\\
& x^3=R_2 \cos\varphi_2, \,\, x^4= R_2 \sin\varphi_2,\nonumber\\
&x^5=x^6=0.
\end{align}
and the coupling to the scalars are still given by~\eqref{scalars}
\begin{equation}
    n^1=-\sin (\varphi_1+\varphi_2), \,\, n^2=\cos (\varphi_1+\varphi_2).
\end{equation}
Choose the worldvolume
coordinates as $\{\sigma^1, \sigma^2, \sigma^3\}=\{\varphi_2, v,z\}$, the dual membrane  solution is~\cite{Drukker:2021vyx},
\begin{align}\label{eq:cylinder}
    x^1&=v\,,\,\,
    x^2=x^5=x^6=0\,,\nonumber\\
    x^3&=r_2(z) \cos \varphi_2\,,\,\, x^4=r_2(z) \sin \varphi_2\,,\nonumber\\
    r_2(z)&=\sqrt{R^2_2-\frac{z^2}2},\nonumber\\
    \phi&=\varphi_2\,,\,\, 
    \cos\theta=\frac{z^2}{2R_2^2}\,,\nonumber
    \\\alpha&=\alpha_0\,,\,\, \beta=\beta_0\,.
\end{align}
Then the on-shell action of an individual
probe membrane taking into account the Legendre transformation is 
\begin{equation}
    S_{\text{ren.}}^{\text{cyl.}}=-\frac{ND}{2R_2}  \,,
\end{equation}
with $D=2\pi R_1$ serving as the IR cutoff for the length of the cylinder.
As the toroidal case, we have, 
\begin{equation}
    \begin{split}
         \langle W(S) \rangle& =\frac{1}{\text{Vol}(S^2)}\int d\mu (S^2) \exp (- S^{\text{cyl.}}_{\text{ren.}})\\
    &=\exp (-S^{\text{cyl.}}_{\text{ren.}})\\
    &=\exp \left(\frac{ND}{2R_2}\right),
    \end{split}
\end{equation}
since $S^{\text{cyl.}}_{\text{ren.}}$ is independent of $\alpha_0, \beta_0$.

\subsection{Supergravity modes} 
In this subsection, we review the fluctuation of the background fields in $AdS_7\times S^4$ dual to the CPO. These fluctuations are referred as supergravity modes or simply supergraviton. 

The fluctuated metric could be decomposed as
\begin{equation}
    G_{mn}=g_{mn}+h_{mn},
\end{equation}
where $g_{mn}$ is the background metric, $h_{mn}$ is the fluctuations. 

The fluctuation of the three-form gauge potential is 
\begin{equation}
    \delta C_{mnp}=a_{mnp}.
\end{equation}
The supergravity modes corresponding to $\mathcal{O}_\Delta$ is~\cite{Aharony:1998rm, Minwalla:1998rp, Leigh:1998kt, Halyo:1998mc}
\begin{align}
    &h_{{\alpha}{\beta}}=\frac{1}{4}g_{{\alpha}{\beta}}s^{I}Y^I,  \\ &h_{{\mu\nu}}=\frac{3}{16k(2k+1)}\nabla_{({\mu}}\nabla_{{\nu})}s^{I}Y^I-\frac{1}{14}g_{{\mu\nu}}s^{I}Y^I,\\&\delta C_{{\alpha\beta\gamma}}=\sum_{I}\frac{3}{16k}\epsilon_{{\alpha\beta\gamma\delta}}s^{I}\nabla^{{\delta}}Y^{I}.
\end{align}
Here the round bracket in $(\mu\nu)$ means  taking the symmetric traceless part:
\begin{equation}
     \nabla_{( \mu}\nabla_{\nu)}=\nabla_{\mu}\nabla_{\nu}-\frac{1}{7}g_{\mu \nu}\nabla^2,
\end{equation} 
$Y^I$ are spherical harmonic functions on $S^4$, and
$s^I$  is related to the source $s^I_0$ of the chiral primary operator  as  $s^I(\vec{x},z)=\int d^6\vec{x}^{\prime}G_\Delta(\vec{x}^{\prime};\vec{x},z)s_0^I(\vec{x}^{\prime})$, where \begin{equation}
    G_\Delta(\vec{x}^{\prime};\vec{x},z)=c\left(\frac{z}{z^2+|\vec{x}-\vec{x}^{\prime}|^2}\right)^\Delta
\end{equation} is the bulk-to-boundary propagator and $c$ is constant for normalization of the boundary-to-bulk propagator:
\begin{equation}
    c=\frac{8^{2+k}(2k-3)(2k-1)(2k+1)\Gamma(k+3/2)}{9\pi^{1/2}N^3\Gamma(k)},
\end{equation}
where we take $\Delta=2k$. We will always apply it in the following, because we only consider the local operator $\mathcal{O}_\Delta$.

\subsection{Holographic Wilson-surface one-point functions}
Then we can compute the Wilson surface one-point functions holographically by
\begin{align}
    \frac{\langle W(\mathcal{S})\mathcal{O}_\Delta\rangle}{\langle W(\mathcal{S})\rangle}\sim-\frac{1}{\mathcal{N}^I}\frac1{\mathrm{Vol}(S^2)}\int \mu(S^2)\frac{\delta S_{M2}}{\delta s_0^I(\vec{x})},
\end{align}
with
\begin{equation}
    \mathcal{N}^I=-2^{3k/2+3}\frac{(2k-3)(2k+1)}{3\pi^{1/4}N^{3/2}}\sqrt{\frac{(2k-1)\Gamma(k+1/2)}{\Gamma(k)}}.
\end{equation}
This normalization factor is fixed by making   the
coefficient of the 2-point function  unit~\cite{Corrado:1999pi}.


In order to compute the Wilson surface one-point functions, We study $\delta S_{M2}$ from~\eqref{eq:m2action}:
\begin{equation}
\delta S_{\text{M2}} = \frac{1}{2} T_{\text{{M2}}} \int d\mathrm{Vol} \, g^{ab} h_{ab}-T_{\text{M2}}\int P[\delta C_3],
\end{equation}
where $g_{ab}$ is the induced metric of worldvolume and $h_{ab}$ is the corresponding induced fluctuations. 

The following task is to compute the fluctuations $h_{mn}$, or more precisely to focus on $h_{\mu \nu}$ in detail. We consider  the toroidal surface with radii $R_1, R_2$ and local operator at distance $L$ from the origin. The OPE limit is defined through $L\gg R_i, i=1, 2$. In this limit $h_{\mu \nu}$ can be easily obtained as~\cite{Corrado:1999pi}:
\begin{equation}
    h_{{\mu\nu}}\simeq-\frac{1}{8}g_{{\mu\nu}}s^I Y^I+\frac{3}{8}\delta_{{\mu}}^z\delta_{{\nu}}^z\frac{1}{z^2}s^I Y^I.\label{coff in OPE}
\end{equation}
In this work, we are not restricted  to the OPE limit and consider the situation with generic $L, R_i$.
Now the computation of $h_{\mu\nu}$ is more  involved. The first-order partial derivative of the bulk-to-boundary propagator $G_\Delta$ is:
\begin{equation}
     \partial_\mu G_\Delta= \delta_\mu^z \frac{\Delta}{z} G_\Delta-\Delta c^{-\frac{1}{\Delta}}\frac{2}{z}A_\mu G_\Delta^{1+\frac{1}{\Delta}},
\end{equation}
with $ A_\mu:=\delta_\mu^zz+\delta_\mu^i(x^i-x^{\prime i})$. We can obtain $\partial_\nu\partial_\mu G_\Delta$ using $\partial_\mu G_\Delta$. Through complicated computations, we can get
\begin{multline}
         \partial_\nu\partial_\mu G_\Delta=\Delta(\Delta-1)\frac{1}{z^2}\delta_\mu^z\delta_\nu^zG_\Delta-\Delta c^{-\frac{1}{\Delta}}\frac{2}{z^2}(\Delta A_\nu \delta_\mu^z+\Delta A_\mu \delta_\nu^z+z\delta_{\mu\nu})G_{\Delta}^{1+\frac{1}{\Delta}} \\+4\Delta(1+\Delta)c^{-\frac{2}{\Delta}} \frac{1}{z^2} A_\mu A_\nu G_\Delta^{1+\frac{2}{\Delta}}.
\end{multline}
Then we need $
    \nabla_\mu \nabla_\nu G_\Delta=\partial_\mu\partial_\nu G_\Delta-\Gamma^\rho_{\mu \nu}\partial_\rho G_\Delta,
$
and
$\nabla^2 G_\Delta=g^{\mu \nu}\nabla_\mu \nabla_\nu G_\Delta
$. During the computation of them, we perform the following computation:
\begin{multline}
     g^{\mu \nu }\partial_\mu \partial_\nu G_\Delta= \Delta(\Delta-1)G_\Delta-2\Delta c^{-\frac{1}{\Delta}}(2\Delta A_z +7z)G_{\Delta}^{1+\frac{1}{\Delta}} \\+4\Delta(1+\Delta)c^{-\frac{2}{\Delta}} \frac{1}{z^2} A^2 G_\Delta^{1+\frac{2}{\Delta}}, 
\end{multline}
with $A^2:=A_\mu A_\nu g^{\mu \nu}=z^2(z^{2} + \left|\vec{x} - \vec{x}^{\prime}\right|^{2})$, and
\begin{equation}
    g^{\mu \nu }\Gamma^\rho_{\mu \nu}\partial_\rho G_\Delta=z^2\Gamma^\rho_{\lambda\lambda}\partial_\rho G_\Delta,
\end{equation}
where the repeated indices $\lambda$ in $\Gamma_{\lambda\lambda}^\rho$ follow the Einstein summation convention.
We could plug all of them into the expression of $h_{\mu \nu}$ and write the final result in three parts: $G_\Delta$, $G_\Delta^{1+\frac{1}{\Delta}}$, and $G_\Delta^{1+\frac{2}{\Delta}}$.

The coefficient of $G_\Delta$ is
\begin{equation}
    \frac{3}{16k(2k+1)}[\Delta(\Delta-1)\frac{1}{z^2}\delta_\mu^z\delta_\nu^z-\Gamma^z_{\mu \nu}\frac{\Delta}{z}-\frac{1}{7}\Delta(\Delta-1)g_{\mu \nu}+\frac{1}{7}g_{\mu \nu}z\Gamma^z_{\lambda \lambda}\Delta]-\frac{1}{14}g_{\mu \nu}.
\end{equation}
The coefficient of $G_\Delta^{1+\frac{1}{\Delta}}$ is 
\begin{multline}
      \frac{3}{16k(2k+1)}[-2\Delta c^{-\frac{1}{\Delta}}\frac{1}{z^2}(\Delta A_\nu \delta_\mu^z+\Delta A_\mu \delta_\nu^z)+\Delta c^{-\frac{1}{\Delta}}\frac{2}{z}\Gamma^\rho_{\mu \nu} A_\rho\\+\frac{4}{7}\Delta^2 c^{-\frac{1}{\Delta}}g_{\mu \nu} A_z -\frac{2}{7} \Delta c^{-\frac{1}{\Delta}}z g_{\mu \nu }\Gamma^\rho_{\lambda \lambda} A_\rho].
\end{multline}
The coefficient of $G_\Delta^{1+\frac{2}{\Delta}}$ is 
\begin{equation}
    \frac{3}{16k(2k+1)}[4\Delta(1+\Delta)c^{-\frac{2}{\Delta}}\frac{1}{z^2}A_\mu A_\nu -\frac{4}{7}\Delta(\Delta+1)c^{-\frac{2}{\Delta}} \frac{1}{z^2}A^2g_{\mu \nu}].
\end{equation}
At this stage, we can introduce simplifications only for the coefficient of $G_\Delta$. After applying equations :
\begin{align}
    \Gamma^z_{\lambda \lambda}&=\frac{5}{z},\\
     \Gamma_{{\mu\nu}}^{z}&=zg_{{\mu\nu}}-\frac{2}{z}\delta_{{\mu}}^{z}\delta_{{\nu}}^{z}, 
\end{align}
the coefficient of $G_\Delta$ is simplified as
\begin{equation}
    \frac{3}{8}\frac{1}{z^2}\delta_\mu^z\delta_\nu^z-\frac{1}{8}g_{\mu \nu},
\end{equation}
which is the same as the coefficient~\eqref{coff in OPE} in the OPE limit. To further simplify the coefficients of $G_\Delta^{1+\frac{1}{\Delta}}$ and $G_\Delta^{1+\frac{2}{\Delta}}$, we next study:
\begin{equation}
    g^{ab}h_{ab}=g^{ab}\frac{\partial x^m}{\partial \sigma^a}\frac{\partial x^n}{\partial \sigma^b}h_{mn},
\end{equation}
where $x^m,x^n$ refer to the coordinates of the membrane solutions in the $AdS_7\times S^4$ and $\sigma^a,\sigma^b$ refer to the coordinates of the worldvolume. Now in the computation of $g^{ab}h_{ab}$, the coefficient of $G_\Delta$ could be written as $N_1$:
\begin{equation}
    N_{1}=\frac{3}{8}\frac{1}{z^2}g^{zz}-\frac{1}{8}D_A+\frac{1}{4}D_S,
\end{equation}
where we define:
\begin{align}
     D_A&:=g^{ab}\frac{\partial x^\mu}{\partial \sigma^a}\frac{\partial x^\nu}{\partial \sigma^b}g_{\mu \nu},\\
    D_S&:=g^{ab}\frac{\partial x^\alpha}{\partial \sigma^a}\frac{\partial x^\beta}{\partial \sigma^b}g_{\alpha \beta}.
\end{align}
For the coefficient of $G_\Delta^{1+\frac{1}{\Delta}}$, we first compute:
\begin{equation}
\begin{split}
    \Gamma^\rho_{\mu \nu}A_{\rho}\frac{\partial x^\mu}{\partial \sigma^a}\frac{\partial x^\nu}{\partial \sigma^b}g^{ab}&=\Gamma^z_{\mu \nu}A_{z}\frac{\partial x^\mu}{\partial \sigma^a}\frac{\partial x^\nu}{\partial \sigma^b}g^{ab}+\Gamma^i_{\mu \nu}A_{i}\frac{\partial x^\mu}{\partial \sigma^a}\frac{\partial x^\nu}{\partial \sigma^b}g^{ab}\\&=(zg_{{\mu\nu}}-\frac{2}{z}\delta_{{\mu}}^{z}\delta_{{\nu}}^{z})z\frac{\partial x^\mu}{\partial \sigma^a}\frac{\partial x^\nu}{\partial \sigma^b}g^{ab}-\frac{2}{z}g^{zb}\frac{\partial x^i}{\partial \sigma^b}A_{i}\\
    &=z^2D_A-2g^{zz}-\frac{2}{z}g^{zb}\frac{\partial x^i}{\partial \sigma^b}A_{i}\\&=z^2D_A-2g^{zz}-\frac{2}{z}g^{zb}(\frac{\partial x^\mu}{\partial\sigma^b}A_\mu-\frac{\partial z}{\partial \sigma^b}A_z)
    \\& =z^2D_A-\frac{2}{z}B,
\end{split}
\end{equation}
where  $\Gamma^i_{zj}=-\frac{1}{z}\delta^i_j$ with $i, j=1,2...6$ has been used  and  we have defined:
\begin{equation}
B:=g^{ab} A_\mu \delta_\nu^z \frac{\partial x^\mu}{\partial \sigma^a}\frac{\partial x^\nu}{\partial \sigma^b}=g^{zb}\frac{\partial x^\mu}{ \partial\sigma^b}A_{\mu}.
\end{equation}
We use this and $\Gamma^\rho_{\lambda \lambda} A_\rho=\Gamma^z_{\lambda \lambda} A_z=5$, then the coefficient of $G_\Delta^{1+\frac{1}{\Delta}}$ is simplified as
\begin{equation}
   \frac{3}{2}c^{-\frac{1}{\Delta}}(-\frac{B}{z^2}+\frac{1}{7}zD_A).
\end{equation}
Then we could simplify the coefficient of $G_\Delta^{1+\frac{2}{\Delta}}$:
\begin{equation}
   \frac{3}{2}c^{-\frac{2}{\Delta}}(\frac{\tilde{C}}{z^2}-\frac{1}{7}\frac{A^2}{z^2}D_A),
\end{equation}
where we define:
\begin{equation}
    \tilde{C}:=A_\mu A_\nu \frac{\partial x^\mu}{\partial \sigma^a}\frac{\partial x^\nu}{\partial \sigma^b}g^{ab}.
\end{equation}
It turns out that the sum of parts involving $G_\Delta^{1+\frac{1}{\Delta}}$ and $G_\Delta^{1+\frac{2}{\Delta}}$ can be simplified:
\begin{equation}
\begin{split}
    & \frac{3}{2}c^{-\frac{1}{\Delta}}(-\frac{B}{z^2}+\frac{1}{7}zD_A)G_\Delta^{1+\frac{1}{\Delta}}+\frac{3}{2}c^{-\frac{2}{\Delta}}(\frac{\tilde{C}}{z^2}-\frac{1}{7}\frac{A^2}{z^2}D_A)G_\Delta^{1+\frac{2}{\Delta}}\\=&\frac{3}{2}c^{-\frac{2}{\Delta}}\frac{\tilde{C}}{z^2}G_{\Delta}^{1+\frac{2}{\Delta}}-\frac{3}{2}c^{-\frac{1}{\Delta}}\frac{B}{z^2}G_{\Delta}^{1+\frac{1}{\Delta}}.
\end{split}
\end{equation}
For convenience, we could let ${N}_{1+\frac{1}{\Delta}}=-\frac{3}{2}c^{-\frac{1}{\Delta}}\frac{B}{z^2}$ and ${N}_{1+\frac{2}{\Delta}}=\frac{3}{2}c^{-\frac{2}{\Delta}}\frac{\tilde{C}}{z^2}$, then the Wilson surface one-point functions could be written as:
\begin{multline}
\frac{\langle W(\mathcal{S}) \mathcal{O}_{\Delta} \rangle}
     {\langle W(\mathcal{S}) \rangle}
=-\frac{T_{\text{M2}}}{\mathcal{N}^I}\frac1{\mathrm{Vol}(S^2)}\int\mu(S^2)\left(\int\frac12(N_1G_{\Delta}+{N}_{1+\frac{1}{\Delta}}G_\Delta^{1+\frac{1}{\Delta}}\right.\\\left.+N_{1+\frac{2}{\Delta}}G_\Delta^{1+\frac{2}{\Delta}})
Y^{I} d\text{Vol}-\int\frac{P[\delta C_3]}{\delta s_0^I(\vec{x})}\right).  \label{eq:correlators}  
\end{multline}

\subsection{Computation of toroidal surface correlators}
We now compute the Wilson surface one-point function using \eqref{eq:correlators} for the toroidal surfaces whose dual membrane solutions are given in~\eqref{eq:torus}.

We start with the Wess-Zumino part. The variation of the pullback of the bulk three form gauge potential to the worldvolume is 
\begin{equation}P[\delta{C}_3]=\delta C_{{\alpha \beta \gamma}}\frac{\partial x^{{\alpha}}}{\partial\sigma^a}\frac{\partial x^{{\beta}}}{\partial\sigma^b}\frac{\partial x^{{\gamma}}}{\partial\sigma^c}d\sigma^a\wedge d\sigma^b\wedge d\sigma^c,
\end{equation}
where
\begin{equation}
    \delta C_{\alpha\beta\gamma} = \sum_{I} \frac{3}{16k} \epsilon_{\alpha\beta\gamma\delta} s^{I} \nabla^{\delta} Y^{I}.
\end{equation}
It is not hard to see that $\delta P[C_3]$ vanishes for the case at hand.

The induced metric of the worldvolume is:
\begin{align}
    &g_{\varphi_1\varphi_1}=-\frac{1}{4} + \frac{R_1^2}{z^2} - \frac{z^4}{16 R_2^4},\\
    & g_{\varphi_2\varphi_2}=-\frac{1}{4} + \frac{R_2^2}{z^2} - \frac{z^4}{16 R_2^4},\\
    &g_{\varphi_1\varphi_2}=\frac{1}{4} - \frac{z^4}{16 R_2^4},\\
     &g_{zz}=\frac{1}{4 R_1^2 - 2 z^2} + \frac{1}{4 R_2^2 - 2 z^2} -\frac{4 R_2^4}{-4 R_2^4 z^2 + z^6},\\
     &g_{z\varphi_1}=g_{z\varphi_2}=0.
\end{align}
The volume form of the worldvolume is 
\begin{equation}
    \begin{split}
d\text{Vol}_{\text{tor}}=&\sqrt{\det{(g_{\mu\nu})}}d\varphi_1\wedge d\varphi_2 \wedge dz
\\ =&\frac{ 2 R_2^4 z^2 + R_2^2 z^4 + z^6 - R_1^2 (8 R_2^4 + 2 R_2^2 z^2 + z^4)}{
 4 R_2^2 z^3\sqrt{(2 R_1^2 -z^2) (2 R_2^2 + z^2)}}d\varphi_1\wedge d\varphi_2 \wedge dz.
    \end{split}
\end{equation}
From the induced metric, we can have:
\begin{align}
D_S&=\frac{-2R_2^{2}\left(R_1^{2}+R_2^{2}\right)z^{2}+\left(-3R_1^{2}+R_2^{2}\right)z^{4}+2z^{6}}{2R_2^{4}z^{2}+R_2^{2}z^{4}+z^{6}-R_1^{2}\left(8R_2^{4}+2R_2^{2}z^{2}+z^{4}\right)},
\\
D_A&=
\frac{8R_2^{4}z^{2}+2R_2^{2}z^{4}+z^{6}-4R_1^{2}R_2^{2}\left( 6R_2^{2}+z^{2}\right)}
{2R_2^{4}z^{2}+R_2^{2}z^{4}+z^{6}-R_1^{2}\left( 8R_2^{4}+2R_2^{2}z^{2}+z^{4}\right)},
\\g^{zz}&=
\frac{z^{2}\left(-2R_1^{2}+z^{2}\right)\left(-4R_2^{4}+z^{4}\right)}
{-z^{2}\left(2R_2^{4}+R_2^{2}z^{2}+z^{4}\right)
+R_1^{2}\left(8R_2^{4}+2R_2^{2}z^{2}+z^{4}\right)}.
\end{align}
So $ N_{1}=\frac{3}{8}\frac{1}{z^2}g^{zz}-\frac{1}{8}D_A+\frac{1}{4}D_S=0$. This leads to the fact that \textit{the Wilson surface
one-point function vanishes in the OPE limit ($L\gg R_i$)}. In order to compute  ${N}_{1+\frac{1}{\Delta}}, {N}_{1+\frac{2}{\Delta}}$ from $B$ and $\tilde{C}$, we let the coordinates of the local operator become
\begin{equation}
    x^\prime=(y_1\cos{\gamma_1},y_1\sin{\gamma_1,}y_2\cos{\gamma_2},y_2\sin{\gamma_2},y_3\cos{\gamma_3},y_3\sin{\gamma_3)}.
\end{equation}
Here $y_i, (i=1, 2, 3)$ are raduial coordinates of the $x^{2i-1}$-$x^{2i}$ plane. 
It turns out that the integrand are extremely complicated and difficult to integrate when the surface is probed by the local operator at a generic point. So we pick out the two different kinds of placement of the local operator  and obtain analytic solutions and numerical solutions, respectively.

\paragraph{The first kind of placement.}

Recall that the toroidal Wilson surface only has non-trivial profile in the four-dimensional subspace determined by $x^5=x^6=0$. In this first case, we put the CPO at the origin of this space by setting $y_1=y_2=0$.

Such placement eliminates the trigonometric functions of the denominator in the final integrand. This makes it possible to derive analytic solutions.
In this situation, we find $A_\mu\partial_ax^\mu=\partial_a( \sum_\mu x^\mu x^\mu)$, and thus $B$ and $\tilde{C}$ are
\begin{align}
      B&=g^{zb}\frac{\partial x^\mu}{ \partial\sigma^b}A_{\mu}=\frac{1}{2}g^{zb}\frac{\partial \sum _\mu x^\mu x^\mu}{ \partial\sigma^b}=0,\\
     \tilde{C}&=A_\mu A_\nu \frac{\partial x^\mu}{\partial \sigma^a}\frac{\partial x^\nu}{\partial \sigma^b}g^{ab}=\frac{1}{4}\frac{\partial \sum _\mu x^\mu x^\mu}{\partial \sigma^a}\frac{\partial \sum _\nu x^\nu x^\nu}{\partial \sigma^b}g^{ab}=0.
\end{align}
Here, $\sum_{\mu=1}^7 x^\mu x^\mu=z^2+\sum_{i=1}^6 x^i x^i=R_1^2+R_2^2$, is a constant for the toroidal surfaces. So we derive the fact that the Wilson surface
one-point function is zero with this kind of placement of the local operator. Actually, for any Wilson surface with $\sum _\mu x^\mu x^\mu$ a constant (like the toroidal surface and the spherical surface), when they are probed by the local operator which is placed at the origin of the sub-space where the Wilson surfaces live with non-zero coordinates, the conclusion that $B=\tilde{C}=0$ always holds. Notice that we even need not compute the worldvolume induced metric to get the conclusion.

\paragraph{The second kind of placement.}
Now we turn to the case with the local operator placed at the origin of a plane along the $x^5, x^6$ directions, orthogonal to the above $\mathbb{R}^4$ containing the surface operator. The result is quite complicated to obtain the analytic results. We will list some numerical results for the Wilson-surface one-point functions in stead.  In order to conduct a numerical computation, 
We set $k=2$, $R_1 =2r$, $R_2= r$, $\gamma_1=\gamma_2=0$. Notice that we should finally take an average over the moduli space as in~\eqref{eq:correlators}.  Because only the spherical harmonics are possibly relevant with the coordinates $\alpha_0,\beta_0$ in the final integrand, it is equivalent to take an average only of the spherical harmonics:
\begin{equation}
    \bar{Y}^I=\frac{1}{\text{Vol}(S^2)}\int d\mu (S^2) Y^I(\alpha_0,\beta_0). 
\end{equation}
 If we consider $Y^I$ which is invariant under $SO(3)_{345}$ (such normalized spherical harmonics are listed in the Appendix~\ref{harmonics}), we have $\bar{Y}^I=Y^I$.

\begin{figure}[H]
    \centering
    \begin{subfigure}{0.45\textwidth}
        \centering
        \includegraphics[width=\linewidth]{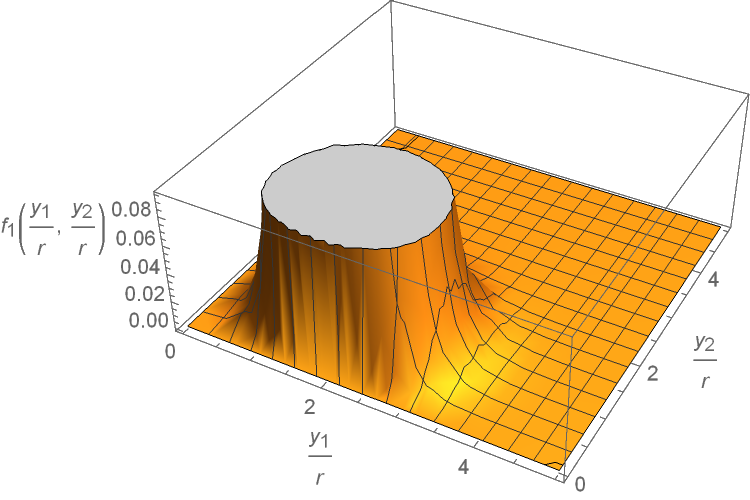} 
        \caption{Three-dimensional figure of $f_1(\frac{y_1}{r},\frac{y_2}{r})$}
        \label{Fig1a}
    \end{subfigure}
    \hfill 
    \begin{subfigure}{0.45\textwidth}
        \centering
        \includegraphics[width=\linewidth]{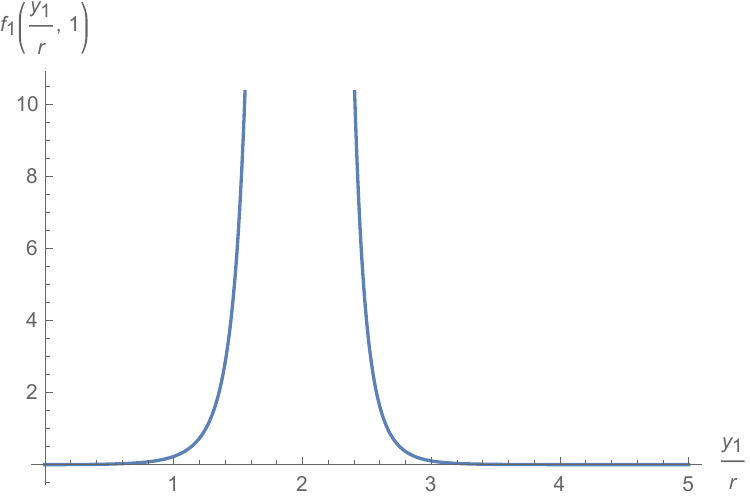} 
         \caption{ $f_1(\frac{y_1}{r},1)$}
         \label{Fig1b}
    \end{subfigure}
    \caption{$f_1(\frac{y_1}{r},\frac{y_2}{r})$}
\end{figure}

Now we choose $Y^I$ as $Y_1^I$ of Appendix~\ref{harmonics}, the Wilson surface
one-point function could be written as:
\begin{equation}
\frac{\langle W(\mathcal{S}_{R_1=2 R_2}) \mathcal{O}_{\Delta}^{Y_1^I} \rangle}
     {\langle W(\mathcal{S}_{R_1=2 R_2}) \rangle}
=\frac{1}{\sqrt{N}r^4}f_1(\frac{y_1}{r},\frac{y_2}{r}).
\end{equation}
We could plot a three-dimensional figure of $f_1(\frac{y_1}{r},\frac{y_2}{r})$ as Figure~\ref{Fig1a} and a two-dimensional figure of $f_1(\frac{y_1}{r},1)$ to show the singularity as Figure~\ref{Fig1b}.
If we select $Y^I$ as $Y_2^I$ of Appendix~\ref{harmonics}, the Wilson
surface one-point function is
\begin{equation}
\frac{\langle W(\mathcal{S}_{R_1=2 R_2}) \mathcal{O}_{\Delta}^{Y_2^I} \rangle}
     {\langle W(\mathcal{S}_{R_1=2 R_2}) \rangle}
=\frac{1}{\sqrt{N}r^4}f_2(\frac{y_1}{r},\frac{y_2}{r}).
\end{equation}
Then we could plot the figures in the same way.
\begin{figure}[H]
    \centering
    \begin{subfigure}{0.45\textwidth}
        \centering
        \includegraphics[width=\linewidth]{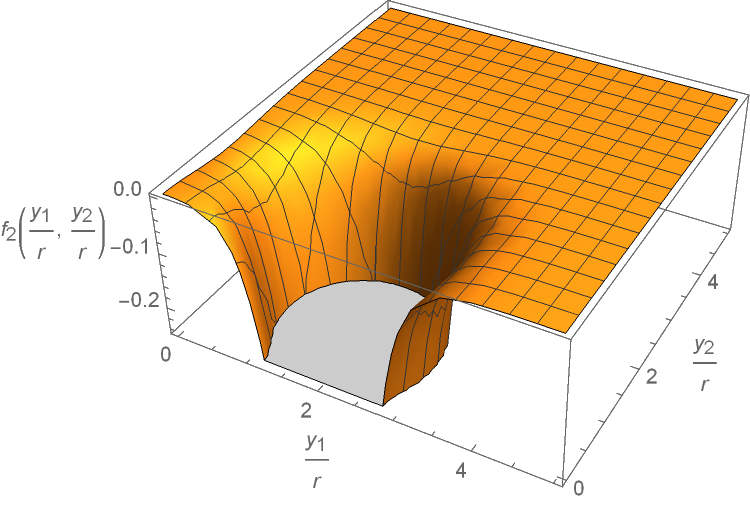} 
        \caption{Three-dimensional figure of $f_2(\frac{y_1}{r},\frac{y_2}{r})$}
    \end{subfigure}
    \hfill 
    \begin{subfigure}{0.45\textwidth}
        \centering
        \includegraphics[width=\linewidth]{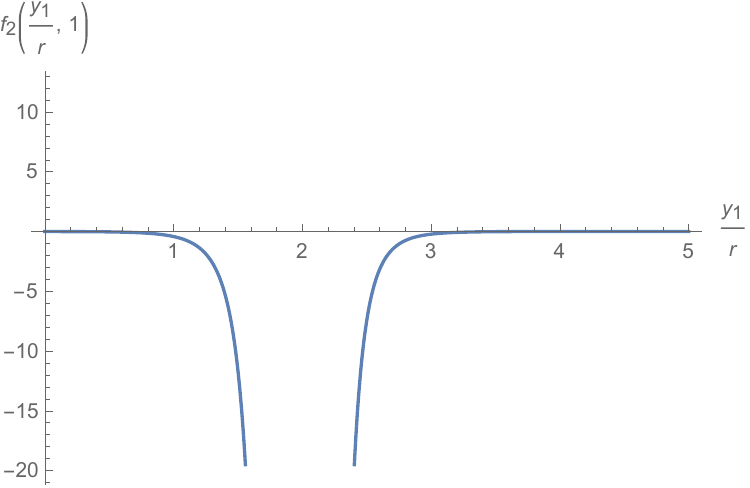} 
         \caption{ $f_2(\frac{y_1}{r},1)$}
    \end{subfigure}
    \caption{ $f_2(\frac{y_1}{r},\frac{y_2}{r})$}
\end{figure}
These figures show the functions $f_i(\frac{y_1}{r},\frac{y_2}{r}), i=1, 2$ are singular at $y_1=2r, y_2=r$, the reason is that the local operator is placed on the Wilson surface. This fact shows the characteristic of the correlation functions. When $y_1=y_2=0$, the placement of the local operator belongs to the first kind at the same time, thus $f_{i}(0,0)=0,i=1,2$.

To show the situation where $R_1=R_2$, we set $k=2$, $R_1 =r$, $R_2= r$, $\gamma_1=\gamma_2=0$, and numerically compute the Wilson surface one-point function by $Y_1^I,Y_2^I$, respectively. We write the Wilson surface one-point function as
\begin{align}
    \frac{\langle W(\mathcal{S}_{R_1= R_2}) \mathcal{O}_{\Delta}^{Y_1^I} \rangle}
     {\langle W(\mathcal{S}_{R_1= R_2}) \rangle}
=&\frac{1}{\sqrt{N}r^4}p_1(\frac{y_1}{r},\frac{y_2}{r}),\\
\frac{\langle W(\mathcal{S}_{R_1= R_2}) \mathcal{O}_{\Delta}^{Y_2^I} \rangle}
     {\langle W(\mathcal{S}_{R_1= R_2}) \rangle}
=&\frac{1}{\sqrt{N}r^4}p_2(\frac{y_1}{r},\frac{y_2}{r}),
\end{align}
and plot figures:
\begin{figure}[H]
    \centering
    \begin{subfigure}{0.45\textwidth}
        \centering
        \includegraphics[width=\linewidth]{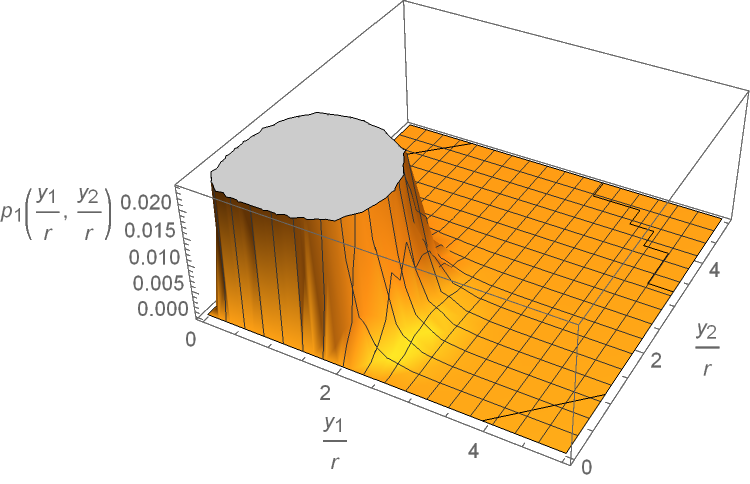} 
        \caption{Three-dimensional figure of $p_1(\frac{y_1}{r},\frac{y_2}{r})$}
    \end{subfigure}
    \hfill 
    \begin{subfigure}{0.45\textwidth}
        \centering
        \includegraphics[width=\linewidth]{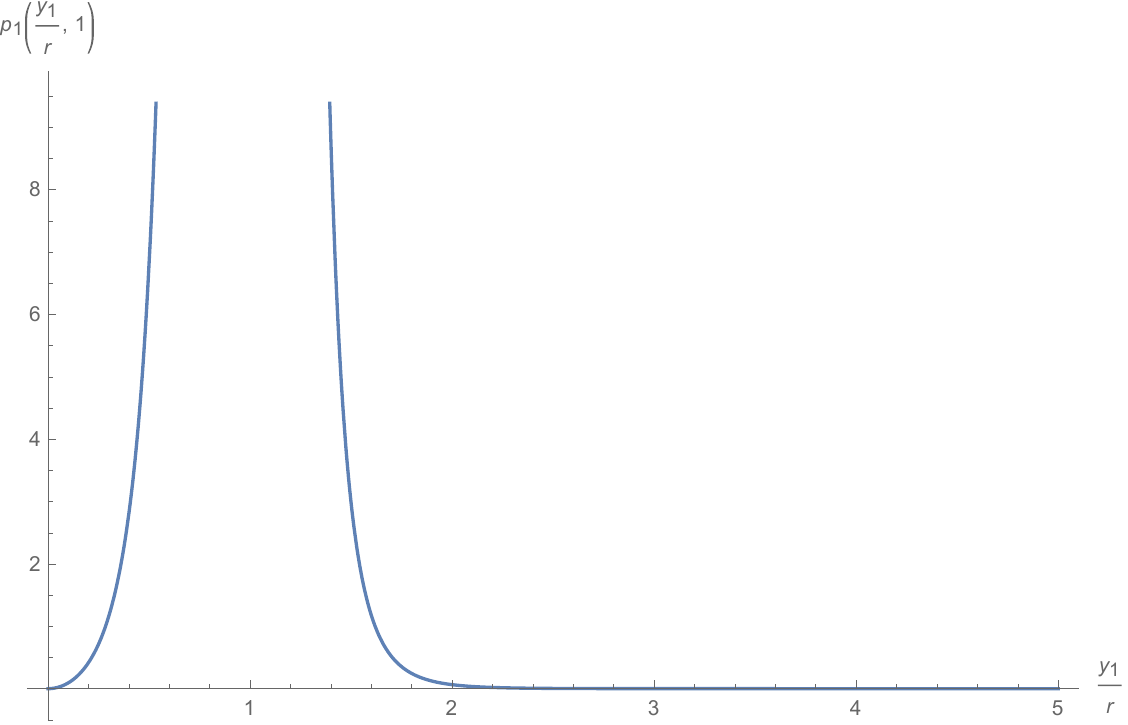} 
    \caption{$p_1(\frac{y_1}{r},1)$}
    \end{subfigure}
    \caption{$p_1(\frac{y_1}{r},\frac{y_2}{r})$}
\end{figure}
\begin{figure}[H]
    \centering
    \begin{subfigure}{0.45\textwidth}
        \centering
        \includegraphics[width=\linewidth]{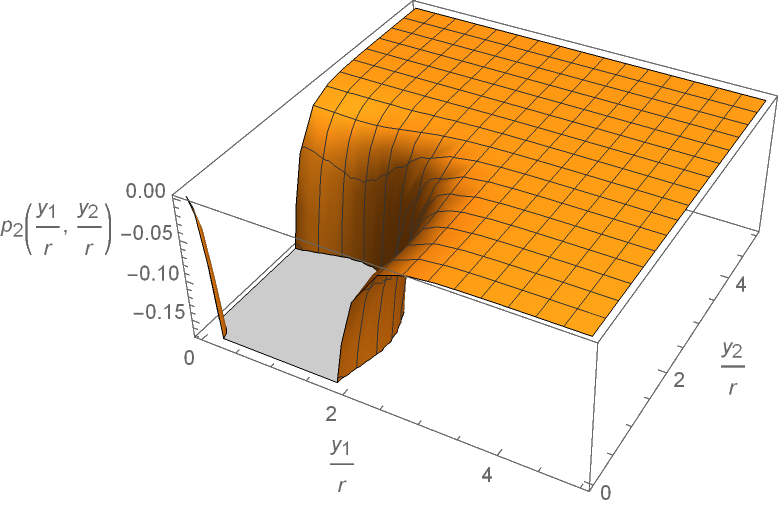} 
        \caption{Three-dimensional figure of $p_2(\frac{y_1}{r},\frac{y_2}{r})$}
    \end{subfigure}
    \hfill 
    \begin{subfigure}{0.45\textwidth}
        \centering
        \includegraphics[width=\linewidth]{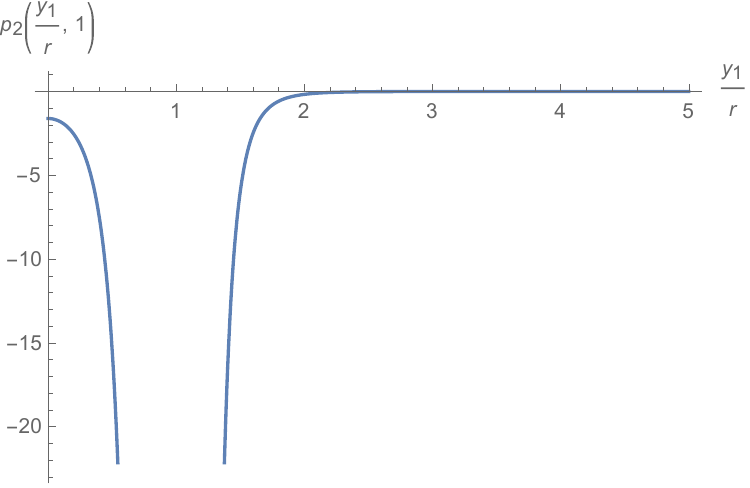} 
         \caption{$p_2(\frac{y_1}{r},1)$}
    \end{subfigure}
    \caption{$p_2(\frac{y_1}{r},\frac{y_2}{r})$}
\end{figure}
Again, we find singularities of $p_i(\frac{y_1}{r},\frac{y_2}{r}),i=1,2$ at $y_1=y_2=r$ and when $y_1=y_2=0$, the placement of the local operator belongs to the first kind as well, causing $p_{i}(0,0)=0,i=1,2$. 

There is an additional feature from $Y_1^I$: When one of $y_1$ or $y_2$ is zero, we obtain $f_1(\frac{y_1}{r},0)=f_1(0,\frac{y_2}{r})=p_1(\frac{y_1}{r},0)=p_1(0,\frac{y_2}{r})=0$ for any $R_1,R_2$, because
we can extract an integral as $\int_0^{2\pi} e^{2i\varphi_1}d\varphi_1=0$ or $\int _0^{2\pi}e^{2i\varphi_2}d\varphi_2=0$ from the final integral in this circumstance.

In order to show the nontrivial influence caused by taking the average over the moduli space, we consider $Y^I$ which is not invariant under $SO(3)_{345}$. We could use such normalized spherical harmonic functions as
\begin{align}
     Y^I_5&=\frac{\sqrt{105}}{8\pi}(\theta^1+i\theta^{3,4,5}),k=2,\\
    Y^I_6&=\frac{\sqrt{105}}{8\pi}(\theta^2+i\theta^{3,4,5}),k=2.
\end{align}
The averages of them over the moduli space are:
\begin{align}
    \bar{Y}^I_5=&-\frac{1}{8\pi}\sqrt{\frac{35}{3}}(\cos^2\theta-3\sin^2 \phi\sin ^2\theta) , \\
     \bar{Y}^I_6=&-\frac{1}{8\pi}\sqrt{\frac{35}{3}}(\cos^2\theta-3\cos^2 \phi\sin ^2\theta) , 
\end{align}
This explicitly shows that the result will change from generic case  due to orbit average. 
 
 We consider the Wilson surface one-point functions in the case of $R_1=R_2$:
\begin{align}
    \frac{\langle W(\mathcal{S}_{R_1= R_2}) \mathcal{O}_{\Delta}^{Y_5^I} \rangle}
     {\langle W(\mathcal{S}_{R_1= R_2}) \rangle}
=&\frac{1}{\sqrt{N}r^4}q_1(\frac{y_1}{r},\frac{y_2}{r}),\\
\frac{\langle W(\mathcal{S}_{R_1= R_2}) \mathcal{O}_{\Delta}^{Y_6^I} \rangle}
     {\langle W(\mathcal{S}_{R_1= R_2}) \rangle}
=&\frac{1}{\sqrt{N}r^4}q_2(\frac{y_1}{r},\frac{y_2}{r}),
\end{align}
Then the figures are:
\begin{figure}[H]
    \centering
    \begin{subfigure}{0.45\textwidth}
        \centering
        \includegraphics[width=\linewidth]{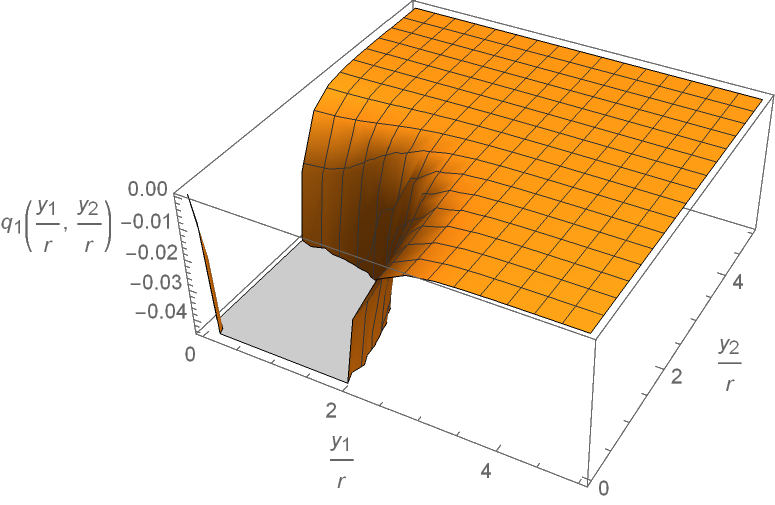} 
        \caption{Three-dimensional figure of $q_1(\frac{y_1}{r},\frac{y_2}{r})$}
    \end{subfigure}
    \hfill 
    \begin{subfigure}{0.45\textwidth}
        \centering
        \includegraphics[width=\linewidth]{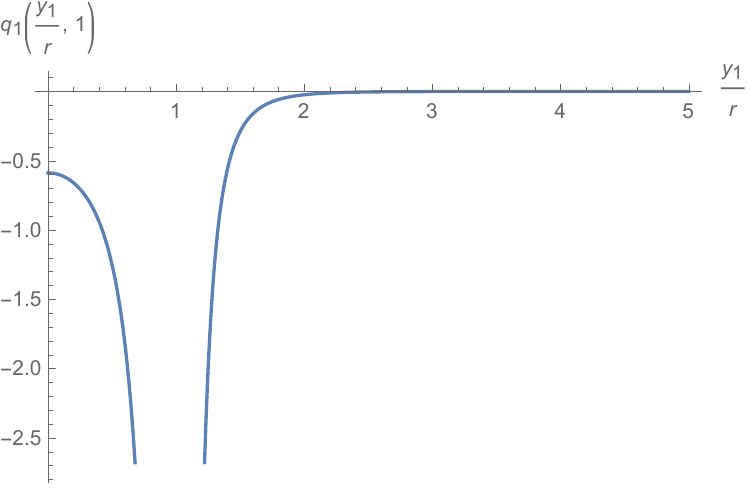} 
    \caption{$q_1(\frac{y_1}{r},1)$}
    \end{subfigure}
    \caption{$q_1(\frac{y_1}{r},\frac{y_2}{r})$}
\end{figure}
\begin{figure}[H]
    \centering
    \begin{subfigure}{0.45\textwidth}
        \centering
        \includegraphics[width=\linewidth]{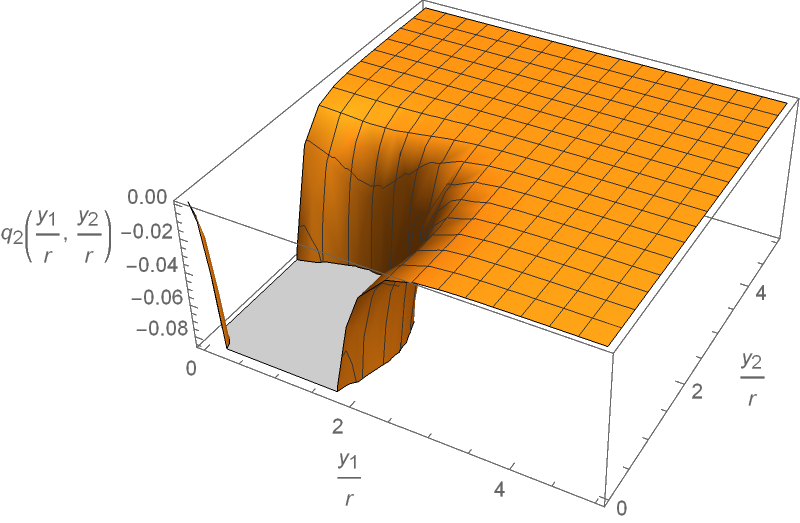} 
        \caption{Three-dimensional figure of $q_2(\frac{y_1}{r},\frac{y_2}{r})$}
    \end{subfigure}
    \hfill 
    \begin{subfigure}{0.45\textwidth}
        \centering
        \includegraphics[width=\linewidth]{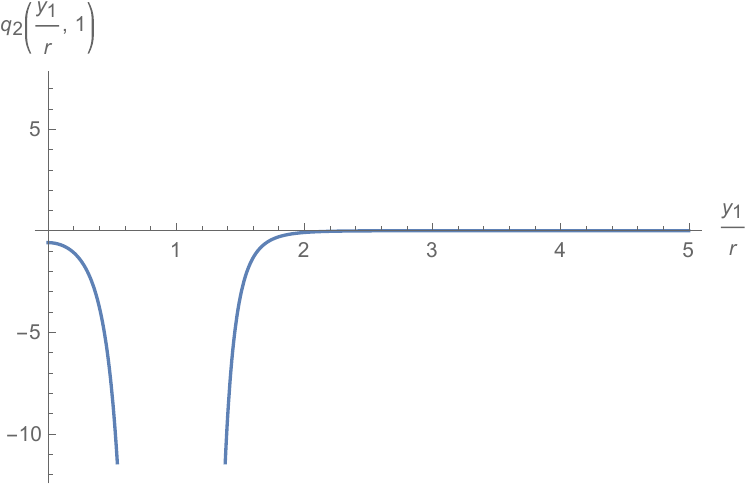} 
         \caption{$q_2(\frac{y_1}{r},1)$}
    \end{subfigure}
    \caption{$q_2(\frac{y_1}{r},\frac{y_2}{r})$}
\end{figure}

\subsection{Computation of cylindrical surface correlators}
\begin{figure}[H]
\centering
\includegraphics[width=.4\textwidth]{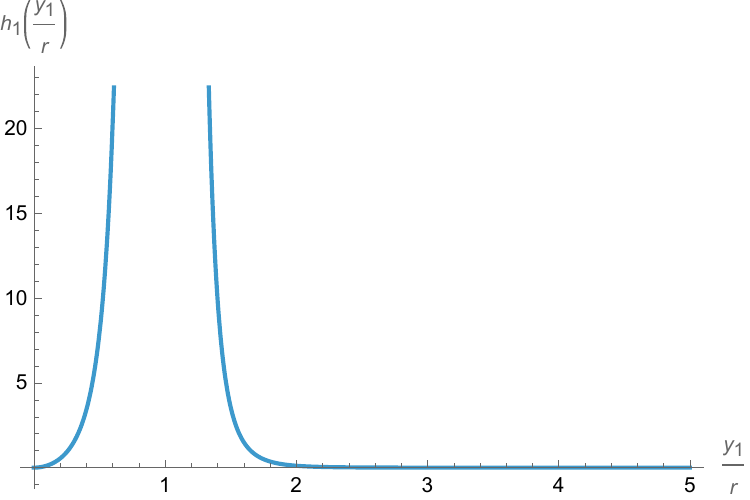}
\caption{$h_1(\frac{y_1}{r})$}
\label{Fig5}
\end{figure}

We now move to the case of cylindrical surface operators with M2-brane dual~\eqref{eq:cylinder}.
We also find that $\delta C_3$ is zero. We then compute $N_1$  from $D_A,D_S$ and the induced metric of the worldvolume. 
The induced metric of the worldvolume is 
\begin{align}
    &g_{\varphi_2\varphi_2}=-\frac{1}{4} +\frac{ R_2^2}{z^2} -\frac{z^4}{16 R_2^4},\\
    &g_{vv}=\frac{1}{z^2},\\
    &g_{zz}=\frac{8 R_2^4 + 2 R_2^2 z^2 + z^4}{8 R_2^4 z^2 - 2 z^6},\\
    &g_{\varphi_2v}=g_{\varphi_2 z}=g_{vz}=0.
\end{align}
The volume form of the worldvolume is 
\begin{equation}
d\text{Vol}_{\text{cyl}}=\sqrt{\det{(g_{\mu\nu})}}d\varphi_2 \wedge dv \wedge dz =\frac{8 R_2^4 + 2 R_2^2 z^2 + z^4}{4 R_2^2 z^3\sqrt{4 R_2^2 + 2z^2}}d\varphi_2 \wedge dv \wedge dz.
\end{equation}
From the induced metric, we can obtain,
\begin{align}
    D_S&=
\frac{2R_2^{2}z^{2}+3z^{4}}{8R_2^{4}+2R_2^{2}z^{2}+z^{4}},\\
    D_A&=
\frac{4R_2^{2}\left(6R_2^{2}+z^{2}\right)}{8R_2^{4}+2R_2^{2}z^{2}+z^{4}}
,\\
 g^{zz}&=
\frac{8 R_2^{4} z^{2} - 2 z^{6}}{8 R_2^{4} + 2 R_2^{2} z^{2} + z^{4}}
.
\end{align}
 So $ N_{1}=\frac{3}{8}\frac{1}{z^2}g^{zz}-\frac{1}{8}D_A+\frac{1}{4}D_S=0$. This time we let the coordinates of the local operator become
\begin{equation}
    x^\prime=(v_0,x^{\prime2},y_1\cos{\gamma_1},y_1\sin{\gamma_1},y_2\cos{\gamma_2},y_2\sin{\gamma_2)}.
\end{equation}
Then we also place the
local operator in two different ways and obtain analytic and numerical solutions, respectively.
\paragraph{The first kind of placement.}
We let $y_1=0$, thus $B$ and $\tilde{C}$ are
\begin{align}
      B= &\frac{4R_2^{4}z^{3}-z^{7}}{8R_2^{4}+2R_2^{2}z^{2}+z^{4}},\\
  \tilde{C}=&4R_2^{4}+R_2^{2}z^2+\left(v-v_0\right)^{2} z^{2}-\frac{z^{4}}{2}-\frac{16R_2^{6}\left(2R_2^{2}+z^{2}\right)}{8R_2^{4}+2R_2^{2}z^{2}+z^{4}}.
\end{align}
If we select $Y^I$ as $Y_1^I$, we can extract an integral as $\int _0^{2\pi}e^{2i\varphi_2}d\varphi_2=0$ from the final integral, so 
\begin{equation}
    \int{N}_{1+\frac{1}{\Delta}}G_\Delta^{1+\frac{1}{\Delta}}
Y_1^{I} d\text{Vol}_{\text{cyl}}=\int{N}_{1+\frac{2}{\Delta}}G_\Delta^{1+\frac{2}{\Delta}}
Y_1^{I} d\text{Vol}_{\text{cyl}}=0,
\end{equation}
which means that the Wilson surface one-point function is zero when $Y^I=Y_1^I$. 
If we select $Y^I$ as other normalized spherical harmonics of Appendix~\ref{harmonics} and let $x^{\prime2}=y_2=0$ in addition, we have:
\begin{table}[htbp]
\centering
\begin{tabular}{c|c|c}
\hline
$Y^I$&k&$\frac{\langle W(\mathcal{S}) \mathcal{O}_{\Delta} \rangle}
     {\langle W(\mathcal{S}) \rangle}$\\
\hline
$Y_2^I$&2 & $\frac{105\sqrt{11}(-2621 + 3780 \log2)}{128\sqrt{2N}R_2^4}$\\
\hline
$Y_3^I$&3 & $-\frac{4851\sqrt{5}(-426203 + 614880 \log2)}{4096\sqrt{N}R_2^6}$\\
\hline
$Y_4^I$&4 & $\frac{2145 \sqrt{95} (-677160409 + 976935960 \log2)}{65536 \sqrt{N} R_2^8}$\\
\hline
$\bar{Y}_{5,6}^I$&2 & $-\frac{35 \sqrt{105}}{512 \sqrt{N} R_2^4}$\\
\hline
\end{tabular}
\caption{Correlation functions with different  spherical harmonics}
\end{table}
\paragraph{The second kind of placement.}
We place the local operator as $y_2=x^{\prime2}=0$, and set $k=2$, $R_2= r$, $\gamma_1=0$ to numerically calculate the Wilson surface one-point function. We first choose $Y^I$ as $Y_1^I$, then the Wilson surface one-point function could be written as:
\begin{equation}
\frac{\langle W(\mathcal{S}) \mathcal{O}_{\Delta}^{Y_1^I} \rangle}
     {\langle W(\mathcal{S}) \rangle}
=\frac{1}{\sqrt{N}r^4}h_1(\frac{y_1}{r}).
\end{equation}
We could plot Figure~\ref{Fig5} along the radial direction of the cylinder.
If we select $Y^I$ as $Y_2^I$, the Wilson surface one-point function is
\begin{equation}
\frac{\langle W(\mathcal{S}) \mathcal{O}^{Y_2^I}_{\Delta} \rangle}
     {\langle W(\mathcal{S}) \rangle}
=\frac{1}{\sqrt{N}r^4}h_2(\frac{y_1}{r}),
\end{equation}
and we could plot a figure in the same way:
\begin{figure}[H]
\centering
\includegraphics[width=.4\textwidth]{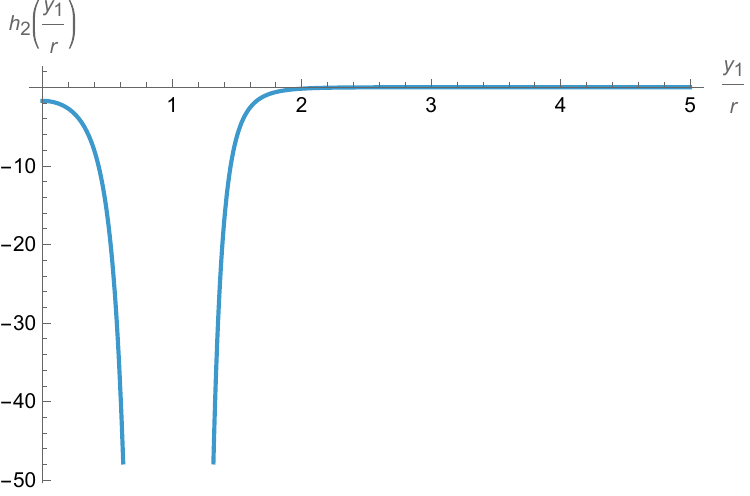}
\caption{$h_2(\frac{y_1}{r})$}
\end{figure}
These figures indicate that $h_{i}(\frac{y_1}{r}), i=1,2$ has a singularity at $y_1=r$, where the local operator is placed on the Wilson surfaces. When $y_1$ is zero, the placement of the local operator belongs to the first kind at the same time, so $h_1(0)=0$ and $h_2(0)=\frac{105\sqrt{11}(-2621 + 3780 \log2)}{128\sqrt{2}}$.

\section{Conclusion}

When a non-local operator is dual to a collection of branes or strings, the computation of the corresponding correlators should include an averaging over this moduli space, which is an orbit of a certain group action.
With this orbit average in mind, we studied the Wilson surface one-point functions in detail when the surface
operator is a $1/8$-BPS toroidal one.  This case is more complicated than the one with a planar or spherical surface since now the dynamical information is also contained in the position dependence of the Wilson-surface one-point functions. We showed that the corerlators vanish  in the OPE limit while the result is non-zero for the generic case. We found that when the CPO is at the origin of the four-dimensional space including the surface operator, the correlation functions vanish. When the CPO is placed at a general place, the correlators are quite complicated and we provide some numerical result when the CPO is put at the origin of the plane orthogonal to the above four-dimensional space. We also investigate the Wilson-surface one-point function for the cylindrical surface operator.

The authors of~\cite{Mori:2014tca} put the $(2, 0)$ theory on $S^1
\times S^5$ and find remarkable agreement on the vev's of Wilson surface between field theory and M-theory sides. In the field theory side, the compactification leads to five-dimensional MSYM on $S^5$ and the vev's were  computed through reduction to the Chern-Simons matrix model via supersymmetric localization. 
In the M-theory side, the vev's are computed from the M5-brane solutions. It is interesting to update this comparison to the Wilson-surface one-point function. On the M-theory side, computations of bulk-to-boundary propagators for CPO's have
already become a challenge due to the twisted periodic boundary condition along the $S^1$ direction~\cite{Kim:2012ava, Kim:2012qf}.

\section*{Acknowledgments}
JW would like to thank N.~Drukker and Z.~Kong for very helpful discussions. 
This work is supported  by the National Natural
Science Foundation of China (NSFC) Grants No.~12375006, 11975164,  12247103, and 
Tianjin University Self-Innovation Fund Extreme Basic Research Project Grant No.~2025XJ21-0007.

\appendix
\section{Normalized spherical harmonics }\label{harmonics}
The spherical harmonics function $Y^I$ is the eigenfunction of the Laplace operator
\begin{equation}
    \nabla^\alpha \nabla_\alpha Y^I=-k(k+3)Y^I.
\end{equation}
Here we could conduct
\begin{equation}
    \nabla^2Y^I=\frac{1}{\sqrt{g}}\partial_\alpha(\sqrt{g}g^{\alpha \beta}\partial_\beta Y^I).
\end{equation}
The solution is as follows 
\begin{equation}
    Y^I=C_{i_1...i_k}\theta^{i_1}...\theta^{i_k},
\end{equation}
where $C_{i_1...i_k}$ is a traceless symmetric tensor:
\begin{equation}
    C_{i_1...i_k}=C_{(i_1...i_k)},\delta^{i_1i_2} C_{i_1...i_k}=0.
\end{equation}
If we consider $Y^I$ which is invariant under $SO(3)_{345}$, we could use the normalized spherical harmonic function as
\begin{equation}
     Y_1^I=\frac{\sqrt{\Gamma(\frac{5}{2}+k)}}{\sqrt{2}\pi^{\frac{5}{4}}\sqrt{\Gamma(1+k)}}(-\sin \theta \sin \phi + i\sin \theta \cos\phi)^k, k\in Z^+
\end{equation}
If we consider $Y^I$ which is invariant under $SO(2)_{12}\times SO(3)_{345}$, the solution should depend only on $\theta$, we get the equation: 
\begin{equation}
    \frac{1}{\sin \theta \cos^2 \theta}\partial_\theta(\sin \theta \cos^2 \theta Y^I)=-k(k+3)Y^I.
\end{equation}
The solutions that are smooth at $\theta \in [0,\pi/2]$ are
\begin{equation}
    Y^I=\frac{P_{2k+1}(\cos \theta)}{\cos \theta},
\end{equation}
where $P_l(x)$ is the Legendre polynomial.
In this case, we list some normalized spherical harmonics used in the article:
\begin{align}
    Y_2^I&=\frac{1}{16\pi}\sqrt{\frac{11}{2}}(63\cos ^4\theta  - 70\cos^2\theta + 15),k=2,\\
    Y_3^I&=\frac{1}{32\pi}\sqrt{\frac{15}{2}}(429\cos^6\theta - 693\cos^4\theta + 315\cos^2\theta - 35),k=3\\
    Y_4^I&=\frac{1}{256\pi}\sqrt{\frac{19}{2}}[315 - 4620 \cos^2\theta + 
 143 \cos^4\theta (126 - 180 \cos^2\theta + 85 \cos^4\theta)],k=4.
\end{align}

\bibliographystyle{JHEP}
\bibliography{Wu}

\end{document}